\documentclass[twocolumn,numberedappendix,iop]{openjournal}

\usepackage[T1]{fontenc}

\usepackage[bitstream-charter]{mathdesign}
\usepackage{amsmath}
\usepackage{ragged2e}
\usepackage[svgnames]{xcolor}
\usepackage[colorlinks,linkcolor=blue,citecolor=blue,urlcolor=blue]{hyperref}

\usepackage{orcidlink}
\usepackage{graphicx}

\begin{document}

\title{Image formation near hyperbolic umbilic in strong gravitational lensing}

\author{Ashish Kumar Meena\orcidlink{0000-0002-7876-4321}$^\star$}
\thanks{$^\star$\href{mailto:ashishmeena766@gmail.com}{ashishmeena766@gmail.com}}
\affiliation{Physics Department, Ben-Gurion University of the Negev, PO Box 653, Be’er-Sheva 8410501, Israel}
	
\author{Jasjeet Singh Bagla\orcidlink{0000-0002-7749-4155}}
\affiliation{Department of Physical Sciences, IISER Mohali, Sector 81, Knowledge City, Sahibzada Ajit Singh Nagar, Punjab 140306, India\\
National Centre for Radio Astrophysics, Tata Institute of Fundamental Research, Ganeshkhind, Pune 411007, India}
	
\begin{abstract}
Hyperbolic umbilic~(HU) is a point singularity of the gravitational lens equation, giving rise to a ring-shaped image formation made of four highly magnified images, off-centred from the lens centre. Recent observations have revealed new strongly lensed image formations near HU singularities, and many more are expected in ongoing and future observations. Like fold/cusp, image formations near HU also satisfy magnification relation~($R_{\rm hu}$), i.e., the signed magnification sum of the four images equals zero. Here, we study how~$R_{\rm hu}$ deviates from zero as a function of area~($A_{\rm hu}$) covered by the image formation near HU and the distance~($d$) of the central maxima image (which is part of the HU image formation) from the lens centre for ideal single- and double-component cluster-scale lenses. For lens ellipticity values~$\geq0.3$, the central maxima image will form sufficiently far from the lens centre~($d\gtrsim5''$), similar to the observed HU image formations with~$R_{\rm hu}\simeq0$. We also find that, in some cases, double-component and actual cluster-scale lenses can lead to large cross-sections for HU image formations for sources at~$z\gtrsim5$, effectively increasing the chances to observe HU image formation at high redshifts. Finally, we study the time delay distribution in the observed HU image formations, finding that not only are these images highly magnified, but the relative time delay between various pairs of HU characteristic image formation has a typical value of $\sim100$~days, an order of magnitude smaller than generic five-image formations in cluster lenses, making such image formations optimal targets for time delay cosmography studies. 
\end{abstract}

\keywords{Gravitational lensing -- strong; Galaxy cluster -- general}
	
\section{Introduction}
\label{sec:intro}
Strong gravitational lensing, by definition, leads to the formation of multiple images of a background source~\citep[e.g.,][]{1992grle.book.....S}. The exact number and properties of these observed images of a background source depend on the overall geometry of the lens system. For a non-singular lens, the total number of images is always odd~\citep{1981ApJ...244L...1B} and is equal to~$2n+1$, where~$n$ is the number of caustics inside which the source lies. An increase in the complexity of the lens mass distribution results in more complex image formations along with a more complicated caustic structure. An example of this can be seen if we compare the image formation in
galaxy and galaxy cluster lenses. Galaxy lenses typically lead to the formation of double or quad images~\citep[e.g.,][]{2010ARA&A..48...87T} whereas a cluster lens can give rise to relatively much more complex image geometries~\citep[e.g.,][]{2011A&ARv..19...47K, 2024SSRv..220...19N}.

A \textit{generic} case of strong lensing can be defined as where no part of the background source sits on the caustic, leading to isolated images. A source lying on fold/cusp caustics~(also known as stable singularities of the lens mapping) leads to merging images~\citep[also known as \textit{critical} images; e.g.,][]{1986ApJ...310..568B, 1992grle.book.....S} in the form of radial and tangential arcs (with respect to the lens centre). Image formation near point~(or unstable) singularities in strong lensing~\citep[e.g.,][]{1992grle.book.....S, 2001stgl.book.....P} can give rise to complex yet distinct geometries, which can be hard to obtain otherwise~\citep[hence, called ``exotic'' image formations;][]{2021MNRAS.503.2097M, 2021MNRAS.506.1526M, 2022MNRAS.515.4151M, 2023MNRAS.526.3902M}. The most common of these exotic image formations is expected to be the image formation near \textit{swallowtail}~\citep[$A_4$;][]{2001stgl.book.....P} singularity having an arc-like structure made of four images. Image formation near the swallowtail can be obtained if we introduce a localised perturbation~(i.e., a subhalo) near the critical curves, which will lead to a swallowtail-like caustic structure at the source position. Hence, swallowtails are a direct consequence of (order of magnitudes smaller in mass) sub-halos inside the main halo and occur in both galaxy~\citep[e.g.,][]{2002A&A...388..373B, 2004A&A...423..797B} and cluster-scale lenses~\citep[e.g.,][]{1998MNRAS.294..734A} although not every work in literature explicitly states the formation of the swallowtail. That said, the presence of a substructure is not a necessary condition to give rise to a swallowtail singularity. For example, in Figure~4 of \citet{2020MNRAS.492.3294M} and Figure~1 of~\citet{2021MNRAS.503.2097M}, we can see the formation of swallowtails for lenses with similar mass components and far from any galaxy-scale lens components within cluster-scale lenses. Less common than swallowtail (in terms of the observed number of systems) is the image formation near \textit{hyperbolic umbilic}~(HU;~$D_4^+$), which consists of four images in a ring-like shape but off-centred from the lens centre~\citep{2020MNRAS.492.3294M, 2023MNRAS.526.3902M}. HUs are zero shear and unit convergence points in the lens plane~(as discussed in Section~\ref{sec:hu_image}). Thanks to their unique geometry, it is easy to spot image formations near HUs, and they are expected to be more common in cluster lenses as we require non-circular lenses to form HUs. To date, only four to five image formations near HUs have been identified in galaxy clusters~\citep{2008A&A...489...23L, 2023MNRAS.522.1091L, 2023MNRAS.526.3902M, 2024arXiv240411659E} and many more are expected in all-sky surveys~\citep{2021MNRAS.503.2097M, 2021MNRAS.506.1526M}\footnote{Around half a dozen additional HU image formations are claimed to be observed in various cluster lenses~\citep{2021A&A...646A..83R} and it is expected that the details will be published soon.}. Even more rare (than~$A_4$ and~$D_4^+$) is the \textit{elliptic umbilic}~($D_4^-$) due to its extreme sensitivity to the lens parameters and low cross-section for image formation. Image formation near elliptic umbilic is yet to be observed to the best of our knowledge, and a more detailed analysis of such image formations will be presented elsewhere.

Sources lying near the fold and cusp caustics lead to image formation following certain \textit{magnification relations} under which the total signed magnification of fold and cusp images is \textit{zero}~\citep{2003ApJ...598..138K, 2005ApJ...635...35K}. Similarly, image formations near point singularities also satisfy such magnification relations~\citep[but the number of images contributing to the magnification relation is four;][]{2009JMP....50c2501A}. This implies that similar to fold and cusp~\citep[e.g.,][]{1998MNRAS.295..587M}, these image formation near point singularities also have the potential to probe the substructure inside the main lens through deviations in the corresponding magnification relations from the no substructure case~\citep[i.e., deviations from zero;][]{2023MNRAS.526.3902M}. However, similar to fold and cusp, such deviations in magnification can, in principle, occur even in the absence of substructures since mappings near singularities in lensing are only approximations of the ideal singularities occurring in the low-order expansion of Fermat potential~\citep{1986ApJ...310..568B} and the fact that we only observe image formation close to these singularities. A preliminary analysis to determine the level of deviation in HU magnification relation in the absence/presence of subhalos has been presented in~\citet{2023MNRAS.526.3902M} using simulated lenses. Another work,~\citet{2023MNRAS.522.1091L} performed a proof-of-concept analysis to use the astrometric perturbations in HU image formations to detect~$\sim10^9~{\rm M_\odot}$ substructures. In addition to satisfying a magnification relation, HU image formations also have an order of magnitude smaller time delays~($\mathcal{O}(10^2{\rm days})$; similar to what we have in galaxy-scale lenses) compared to generic five image formation in cluster lenses~\citep{2022MNRAS.515.4151M}. Hence, such lensed systems can also be good targets for time-delay cosmography studies\footnote{Although in terms of expected numbers, HU image formations will be less common than galaxy lenses.  However, since such systems primarily occur in cluster lenses, time delay cosmography studies with these systems do not suffer from the well-known mass-sheet degeneracy.  Further, HU image formation constrains the lens model strongly in the vicinity of the images, and hence, there are fewer uncertainties in the lens model.}.

In our current work, motivated by the continuous discoveries of new image formations near HU singularities~\citep{2021A&A...646A..83R, 2023MNRAS.522.1091L, 2023MNRAS.526.3902M, 2024arXiv240411659E} and their ability to detect low-mass substructures~($\lesssim10^9~{\rm M_\odot}$) in the central region of cluster-scale lenses, we focus on the HU image formation and anomalies in the corresponding magnification relation~(or image flux-ratios), extending our previous work presented in~\citet{2023MNRAS.526.3902M}. We study how the magnification relation for image formations near HUs varies as a function of the area covered by the image quadrilateral and the distance of the central maxima image from the lens centre using simple~(single and double component) lenses as well as in real observed HU image formations. We also analyse how the critical HU redshift (i.e., redshift at which HU gets critical) varies in the presence of a second cluster scale lens component, leading us to discover a population of double-component lenses with large cross-sections for image formation near HUs. We also observe similar behaviour in a subset of cluster lenses in the \textit{reionization lensing cluster survey}~\citep[RELICS;][]{2019ApJ...884...85C}. Finally, following~\citet{2022MNRAS.515.4151M}, we study the time delay distribution in the actual observed HU image formations. Since all observed HU image formations are detected in cluster lenses, we also limit ourselves to the cluster-scale lenses.

This work is organised as follows. Section~\ref{sec:hu_image} briefly reviews the properties of HU singularities and the corresponding characteristic image formation. Section~\ref{sec:one_enfw} and~\ref{sec:two_enfw} studies the deviation in magnification relation~($R_{\rm hu}$) as a function of the distance of the central maxima image from the lens centre and area covered by the image formation near the HU. Section~\ref{sec:actual_lens} studies the HU singularities in a sample of actual cluster lenses and image formation properties near them. Section~\ref{sec:obs_hu} studies the properties of observed HU image formations. We conclude our work in Section~\ref{sec:conclusions}. Throughout this work, we use a standard flat LCDM cosmology with parameters~$H_0=70\:{\rm km\:s^{-1}\:Mpc^{-1}}$, $\Omega_{m}=0.3$, and $\Omega_{\Lambda}=0.7$.

\begin{figure*}
    \centering
    \includegraphics[height=17cm, width=17cm]{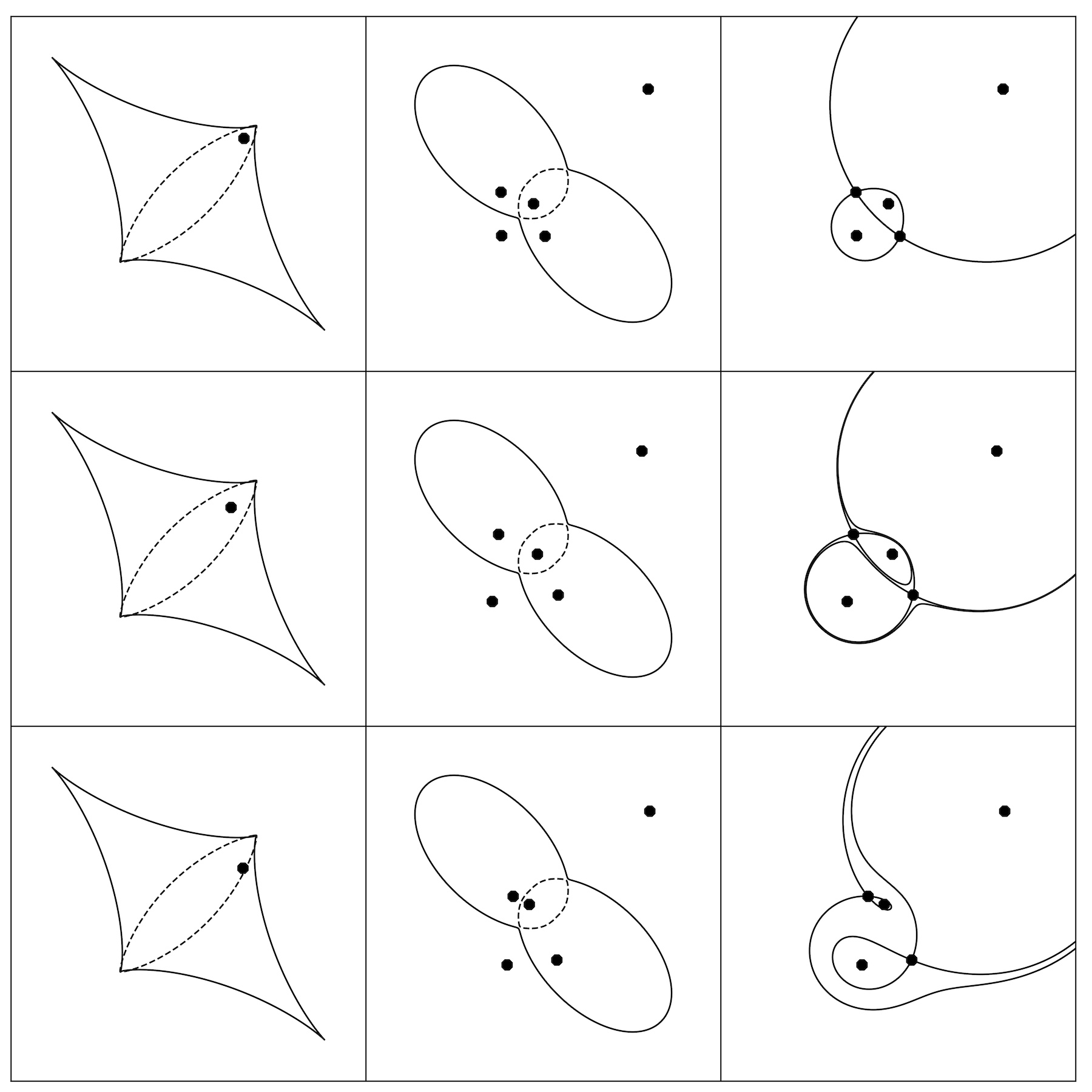}
    \caption{Effect of source position variation on the image formation near a HU singularity for an \textit{elliptical} lens. \textit{Left} column shows the source plane with solid and dashed lines representing the tangential and radial caustics, respectively. The source position is marked by solid black points. \textit{Middle} column represents the corresponding image plane with solid and dashed lines representing the tangential and radial critical curves, respectively, and image positions are given by black solid points. \textit{Right} column shows the corresponding saddle-point time delay contours and image positions.}
    \label{fig:hu_image}
\end{figure*}

\begin{figure*}
    \centering
    \includegraphics[height=10.5cm, width=16cm]{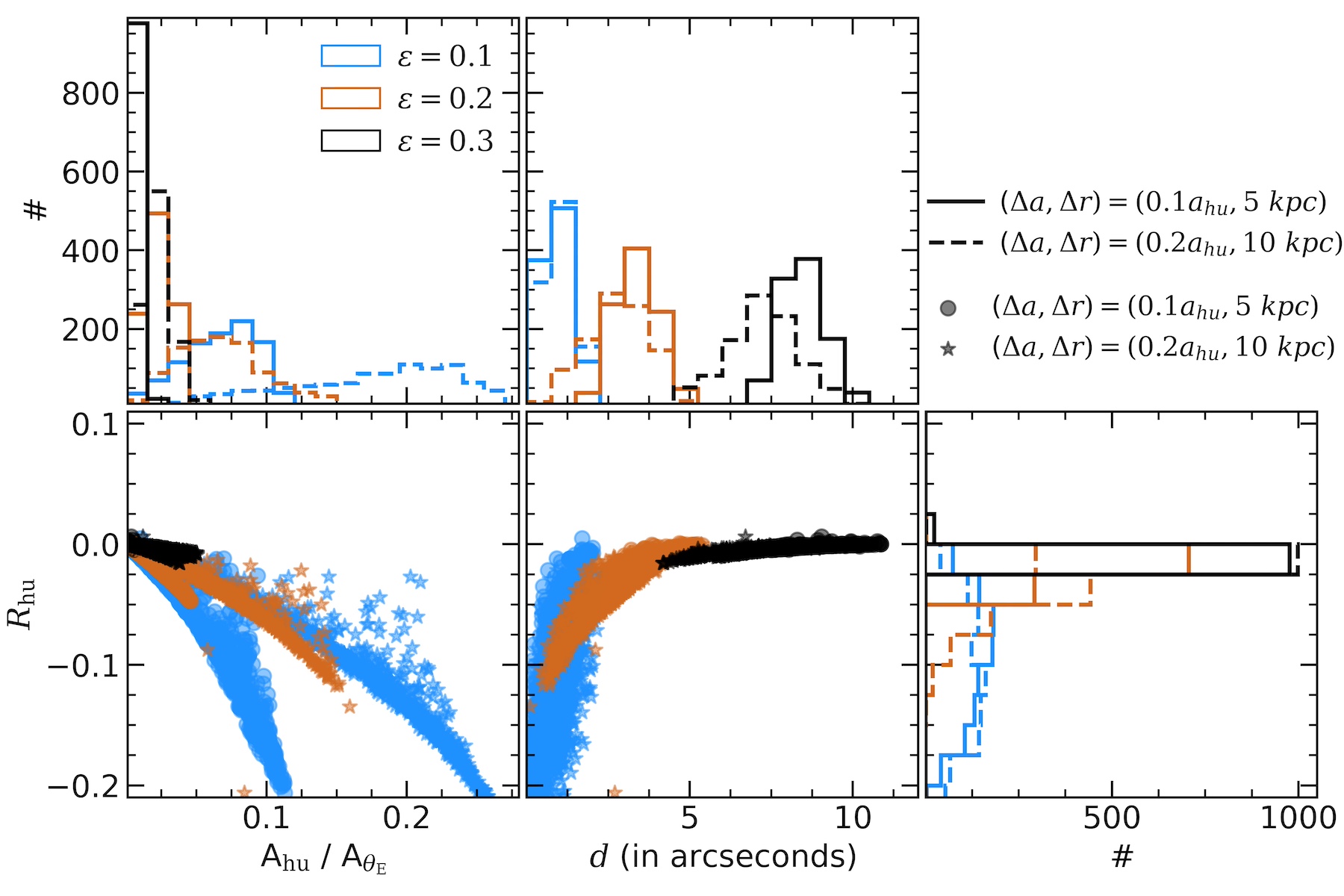}
    \caption{$R_{\rm hu}$ distribution in one-component eNFW lens. The left panel represents the~$R_{\rm hu}$ as a function of the fraction of the area covered by the quadrilateral formed by four images~($A_{\rm hu}$) with respect to the area covered by the Einstein ring~($A_{\theta_E}$). The middle panel shows the~$R_{\rm hu}$ as a function of distance~($d$) of the central maxima image from the lens centre. A small $d$-value implies that the central maxima image forms very close to the lens centre and will not be observable in actual cluster lenses. The histograms show different visualisations of the same information presented in scatter plots. In each panel, the blue, brown, and black points/histograms correspond to ellipticity~($\epsilon$) values of~0.1, 0.2, and 0.3, respectively. For each $\epsilon$-value, we chose two different sets of values for~$(\Delta a, \Delta r)$ to simulate lens systems, where~$\Delta a$ is the variation in distance ratio, and~$\Delta r$ is the radius of the circle centred on the cusp from which the source position was drawn.}
    \label{fig:plot_1enfw_hu}
\end{figure*}

\begin{figure}
    \centering
    \includegraphics[height=10.5cm, width=8cm]{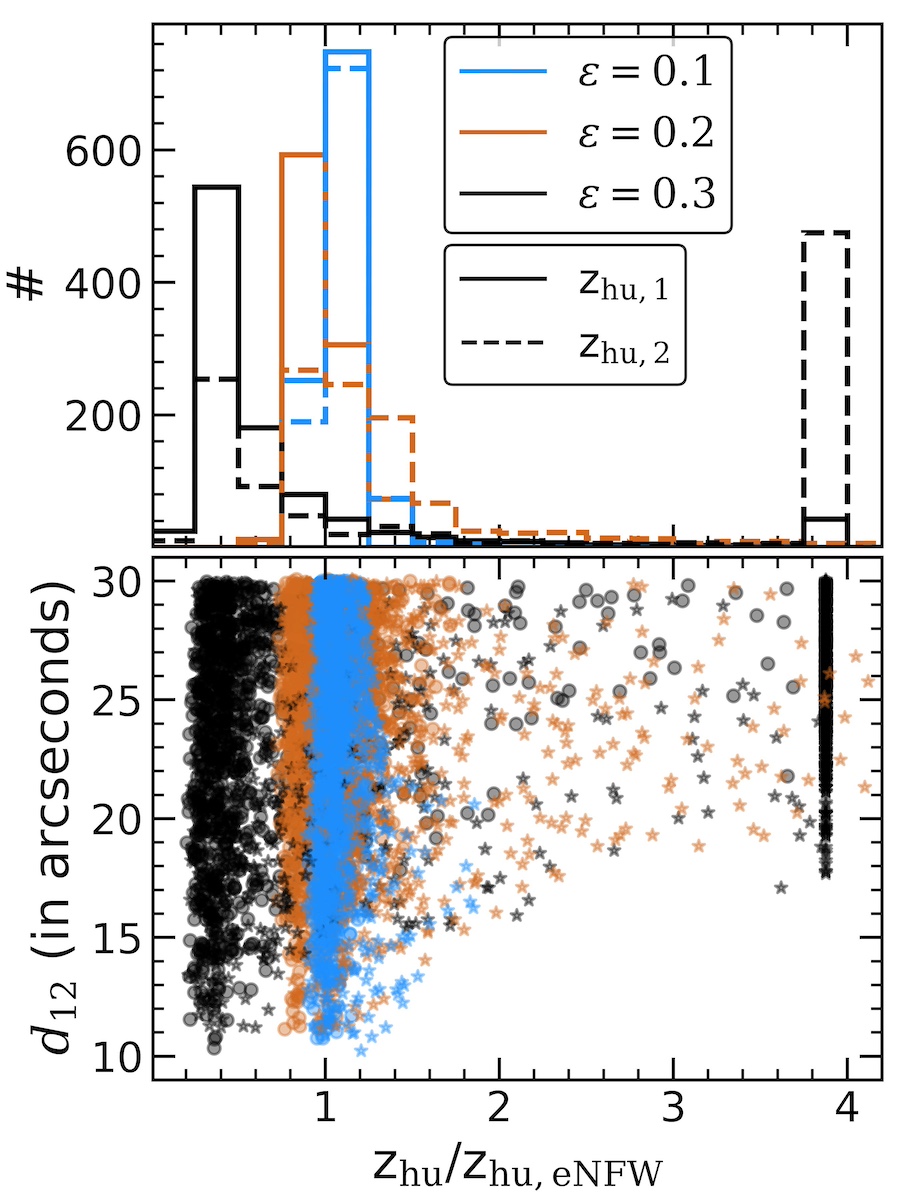}
    \caption{Effect of the presence of second component on~$z_{\rm hu}$. The x-axis represents the ratio of~$z_{\rm hu}$ formed in the presence of a second lens component and~$z_{\rm hu, eNFW}$ without a second component for a given ellipticity~($\epsilon$) value. For~$\epsilon=0.1, 0.2, 0.3$,~$z_{\rm hu, eNFW} = 0.48, 0.66, 2.58$. The y-axis represents the distance~($d_{12}$) between primary and secondary lens components. $z_{\rm hu,1}~(z_{\rm hu,2})$ represents the primary~(secondary) HU points. Primary~(secondary) essentially represents the HU of the first lens component that got critical first~(second).}
    \label{fig:plot_2enfw_zhu}
\end{figure}

\section{HU Image Formation}
\label{sec:hu_image}
The gravitational lens equation relates the observed image position~($\pmb{\theta}$) to the unlensed source position~($\pmb{\beta}$) and given as,
\begin{equation}
    \pmb{\beta} = \pmb{\theta} - \frac{D_{ds}}{D_s}  \pmb{\nabla} \psi (\pmb{\theta}),
    \label{eq:lens}
\end{equation}
where~$\pmb{\nabla}\psi$ represents the deflection vector, which is the gradient of the lensing potential~($\psi$) at the image positions. $a_{\rm dis} \equiv D_{ds}/D_s$ represents the \textit{distance ratio} of angular diameter distances from lens to source and from observer to source. The Jacobian of the lens mapping is given as,
\begin{equation}
    det\mathbb{A} \equiv \mathbb{1} - a_{\rm dis} \psi_{ij} = 
    \mathop{\begin{pmatrix}
        1-a_{\rm dis} (\kappa + \gamma) & 0 \\
        0 & 1-a_{\rm dis} (\kappa - \gamma)
    \end{pmatrix}},
\end{equation}
where~$\kappa$ and~$\gamma$ represent the convergence and shear for a source at infinity. An umbilic singularity, by definition, manifests itself in the image plane where both of the eigenvalues of the deformation tensor~($\psi_{ij}$) are equal to each other~\citep{2020MNRAS.492.3294M}, i.e., 
\begin{equation}
    \kappa+\gamma = \kappa-\gamma.
    \label{eq:umbilic}
\end{equation}
This implies that umbilics are points in the image plane where both eigenvalues of the lensing Jacobian~(i.e.,~$\mathbb{A} \equiv \mathbb{1} - a_{\rm dis} \psi_{ij}$) vanish at the same time, implying that umbilics have~$(a_{\rm dis}\kappa, a_{\rm dis}\gamma)=(1, 0)$. This implies that a point mass lens (or any combination of point mass lenses) cannot lead to umbilic singularities since~$\kappa=0$ everywhere and~$\gamma=0$ points are no longer singular points in the image plane. In the image plane, umbilic singularities mark points where radial and tangential critical curves meet with each other. In the source plane, at the corresponding points, we observe an exchange of cusps between radial and tangential caustics, and the exact number of cusps getting exchanged depends on the type of umbilic singularity. At the hyperbolic umbilic~(HU), we see an exchange of one cusp between radial and tangential caustics, whereas three cusps get exchanged at the elliptic umbilic. Throughout this work,~$z_{\rm hu}$ denotes the source redshift at which an HU gets critical. We can determine the nature of the umbilic singularity from the third-order derivatives of Fermat potential~\citep{1986ApJ...310..568B} in the image plane or first-order derivatives of deformation tensor~($\psi_{ij}$) without going to the source plane and looking at the caustic structure~(see Chapter~6 in~\citealt{1992grle.book.....S} for more details). A more straightforward way is to look at the \textit{singularity map} in which two~$A_3$-lines meet at the HU singularity, and six $A_3$-lines meet at elliptic umbilic~\citep{2020MNRAS.492.3294M} since it gives position of all point singularities along with cusp points. 

The image formation near the HU singularity is unique and consists of four images in a ring-like structure~\citep[e.g.,][]{2020MNRAS.492.3294M} off-centred with respect to the lens centre. Examples of image formations near HU singularity for an elliptical lens are shown in Figure~\ref{fig:hu_image}. The left, middle and right columns show the caustic structure in the source plane, critical curves with image formation in the image plane, and saddle-point time delay contours and image positions in the image plane, respectively. We can observe the exchange of cusp along the minor axis of the lens in the source plane\footnote{For an isolated elliptical lens, due to reflection symmetry, both HU singularities get critical at the same redshift, but this will not happen once we break this symmetry.}. A source lying inside both caustics leads to the formation of five images in the image plane shown in the middle column. The isolated image on the upper-right part of the panel marks the global minima, whereas the rest of the four images lie on the other side of the lens centre. The ring-like image formation shown in the top panel of the middle column can be considered as the \textit{characteristic}/\textit{ideal} HU image formation and, so far, two such image formations are observed~\citep{2008A&A...489...23L, 2023MNRAS.522.1091L}. Image formations shown in the middle and bottom panels are two possible variations of the characteristic HU image formation that we can observe based on varying source positions. Here, we stress that in Figure~\ref{fig:hu_image}, we only varied the source position and fixed the source redshift~($z_s$); however, a complete catalogue of HU image variations should also include variation in~$z_s$~\citep{2022MNRAS.515.4151M}. From the right column, we can see that out of the four images, two are saddle-points, one is minima, and one is maxima. For simple lens models, from the saddle-point contours, we can see that the HU image formation has a lima\c{c}on enclosing a lemniscate~\citep{1986ApJ...310..568B} and that the arrival time order of the four HU images is minima, saddle-points, and maxima, respectively. Out of the two saddle-points, one farther from the maxima (or one on lemniscate) will arrive first. The magnification relation for HU image formation is given as,
\begin{equation}
    R_{\rm hu} \equiv \frac{\mu_1 + \mu_2 + \mu_3 + \mu_4}{|\mu_1| + |\mu_2| + |\mu_3| + |\mu_4|},
\end{equation}
where~$\mu_i$ with~$i=1,2,3,4$ represents the magnifications of the four images (excluding global minima), which are part of the characteristic image formation. In principle, $R_{\rm hu}=0$ is valid only at the HU point~\citep{2009JMP....50c2501A}. However, the same relation can be used for a point in the five image region shown in Figure~\ref{fig:hu_image} as long as the deviations from the HU point lens mapping is negligible. But in the presence of a substructure close to one of the images or if we deviate from the HU lens mapping (by varying the source position or redshift), we can have deviations from zero in~$R_{\rm hu}$~\citep{2023MNRAS.526.3902M}. Hence, deviations from the expected value of zero can be an indication of a departure from the critical position or substructure.

\section{One-component \lowercase{e}NFW lens}
\label{sec:one_enfw}
The simplest lens model that can lead to an HU singularity is an elliptical lens with a central slope shallower than an isothermal profile so that it can give rise to radial caustic since, for HU, we need an exchange of cusp between radial and tangential caustics. In our current work, we use the eNFW profile to study the HU image formation and its properties in ideal cases. For simplicity, we introduce the ellipticity in the potential of a circular NFW profile\footnote{$\epsilon \equiv 1-b/a$; $r^2 \equiv x^2/(1-\epsilon) + (1-\epsilon)y^2$}. Such an approximation will not work for large ellipticities as it will lead to nonphysical shapes for the surface density~\citep{1993ApJ...417..450K, 2023A&A...679A.128G}. However, since we limit ourselves to~$\epsilon\leq0.3$, the above approximation is suited for our work. An eNFW lens can be described by three parameters: total virial mass~($M_{\rm vir}$), concentration parameter~($c_{\rm vir}$), and ellipticity~($\epsilon$). As discussed in~\citet{2023MNRAS.526.3902M}, the critical HU redshift~($z_{\rm hu}$) shows positive correlation with~$\epsilon$ whereas it shows negative correlations with~$M_{\rm vir}$ and~$c_{\rm vir}$. Although in~\citet{2023MNRAS.526.3902M}, it was shown only for an eNFW profile, but we can expect similar behaviour of~$z_{\rm hu}$ with~$\epsilon$ for other lens models.

An isolated elliptical lens, although not representative of actual cluster lenses, allows us to determine how image formations near HU will behave in an ideal case, and in this section, we primarily focus on the effect of~$\epsilon$ on image formation near HU as it is an important parameter in determining the~$z_{\rm hu}$. We fix the other lens parameters to~$(M_{\rm vir}, c_{\rm vir})=(10^{15}{\rm M_\odot}, 7)$ and lens redshift~($z_l$) to~0.3. To study image formation properties near HU, we trace the cusp point around~$z_{\rm hu}$. Figure~\ref{fig:plot_1enfw_hu} shows the effect of lens ellipticity~($\epsilon$) on image formation near HU for two different sets of~$(\Delta a, \Delta r) = (0.1a_{\rm hu}, 5~{\rm kpc})$ and~$(0.2a_{\rm hu}, 10~{\rm kpc})$, where~$\Delta a$ determines the variation in the distance ratio around~$a_{\rm hu}$ (i.e., variation in source redshift) and~$\Delta r$ determines the radius of the circle drawn at the cusp point. Hence~$(\Delta a, \Delta r) = (0.1a_{\rm hu}, 5~{\rm kpc})$ implies that distance the ratio is drawn from the range~$a_{\rm dis} \in [a_{\rm hu}-0.1 a_{\rm hu}, a_{\rm hu}+0.1a_{\rm hu}]$ and source position is drawn from a circle of 5~kpc centred at the cusp point. As discussed in~\citet{2020MNRAS.492.3294M} and~\citet{2023MNRAS.526.3902M}, the HU image formation varies continuously as we change source redshift or position. Hence, in the current work, we vary both of these parameters to study the variation of the HU magnification relation. Only then can we determine the regions around HU singularity, which can be used for flux ratio anomaly studies. The left and middle scatter plots show the magnification relation~($R_{\rm hu}$) as a function of the fraction of the area covered by the four-image quadrilateral compared to the area covered by the effective Einstein ring~($A_{\rm hu}/A_{\rm \theta_E}$)\footnote{Effective Einstein ring area is defined as the area covered by the tangential critical curve.} and~$R_{\rm hu}$ as a function of the distance~($d$) of the maxima image with respect to the centre of the lens. Above~(and in the next section), we choose two different sets of~$(\Delta a, \Delta r)$ for our analysis to see effect of redshift range and source position around~$z_{\rm hu}$ on the~$R_{\rm hu}$.

For~$\epsilon=0.1$, since the lens is more circular compared to other cases, HUs get critical relatively close to the lens centre and at a lower source redshift. Hence, if we draw a circle of 5~kpc, it even covers source positions very close to the lens centre, where the image formation starts to look more like a quad image formation, observed in galaxies, and covers an area of~$\lesssim12\%$ compared to the effective Einstein ring area~(for example, see Figure~\ref{fig:example_1eNFW}). In such cases, maxima will form very close to the lens centre~($\lesssim2.5''$), as we can see from the left panel in Figure~\ref{fig:example_1eNFW}, which can make it very hard to detect as it is de-magnified and will be buried under the light of the brightest cluster galaxy~(BCG) in cluster lenses. Similarly, if we draw a circle of 10~kpc with~$\Delta a = 0.2a_{\rm hu}$, we see that the area covered by the quadrilateral increases and covers a fraction of~$\lesssim30\%$ compared to the Einstein ring area, whereas the maxima image distance have a similar distribution as the 5~kpc case. The effect of having a de-magnified maxima~(forming very close to the lens centre) is significant on utilization of~$R_{\rm hu}$ as we will not be able to detect the maxima image since it will be buried under the light of the lens and on the top of that~$R_{\rm hu}$ shows large deviation (having negative values) from zero. Even though we see a large deviation in~$R_{\rm hu}$ for~$\epsilon=0.1$ case, we can still model it if needed since we see a correlation in the~$R_{\rm hu}$ values with other parameters in Figure~\ref{fig:plot_1enfw_hu}, at least for an isolated elliptical lens since the scatter in~$R_{\rm hu}$ is not huge.

As we move to~$\epsilon=0.2$, for~$(\Delta a, \Delta r) = (0.1a_{\rm hu}, {\rm 5~kpc})$, we see a decrease in the area covered (compared to effective Einstein radius area) by image formation near HU by a factor of two, and the central image position peaks at a distance of~$\simeq3''$ from the lens centre. This implies that for~$\epsilon=0.2$, the image formation near HU will cover less region, so it will actually start to look like the characteristic ring-like image formation and, at the same time, the central image can also be sufficiently far from the lens centre to observe it~(for example, see the middle column in Figure~\ref{fig:example_1eNFW}). The effect of de-magnification of central image on~$R_{\rm hu}$ also starts to subside as nearly all systems show~$R_{\rm hu}\in[-0.05, 0.0]$. For~$(\Delta a, \Delta r) = (0.2a_{\rm hu}, {\rm 10~kpc})$ case, we can see that the distance~($d$) histogram shifts towards lower values and now~$R_{\rm hu}$ has values between -0.05 and -0.1 for ~$\sim40\%$ systems. For very few systems, we see that the~$R_{\rm hu}$ does not follow the overall trend. These are most likely artefacts resulting from our finite resolution; otherwise, we would expect to see more such points. Finally, for~$\epsilon=0.3$, the distance of the central image is always~$\geq4''$ and peaks around~$8''$. At the same time, the area covered by the quadrilateral always remains~$\leq5\%$ of Einstein radius area with~$R_{\rm hu}\simeq0$. Also, see example image formations in the right panel of Figure~\ref{fig:example_1eNFW}.

The above analysis implies that in an isolated cluster-scale lens with~$\epsilon\gtrsim0.3$, we expect to detect all four of the images in image formation near HU with~$R_{\rm hu}\simeq0$ for~$(\Delta a, \Delta r) \leq (0.2a_{\rm hu}, {\rm 10~kpc})$. In addition, the area covered by the four images is also very small compared to the overall strong lensing region (given by the effective Einstein radius area), implying that identification of the image formations near HUs will not be an issue for the above~$(\Delta a, \Delta r)$ values. This does not mean that image formations with~$(\Delta a, \Delta r) > (0.2a_{\rm hu}, {\rm 10~kpc})$ cannot be considered as image formations near HU. Since the image formation will evolve continuously, we need to perform a similar analysis to determine the variation in~$R_{\rm hu}$ for other values of~$(\Delta a, \Delta r)$. However, the determination of lower limits on~$(\Delta a, \Delta r)$ values allows us to estimate the lower limit on the number of image formations near HUs in the large sky surveys, which can also potentially be used to do further flux-ratio anomaly studies.

\begin{figure}
    \centering
    \includegraphics[height=7cm, width=8.5cm]{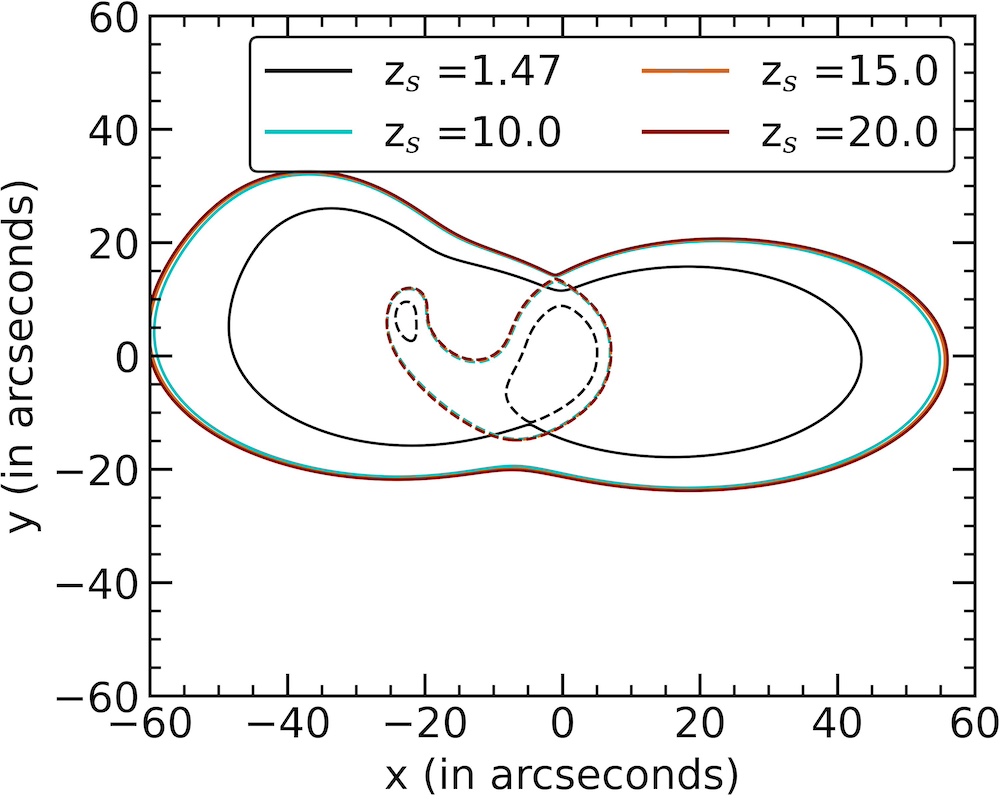}
    \caption{An example of the evolution of critical curves in the presence of a second lens component for~$\epsilon=0.3$ where~$z_{\rm hu,2}$ does not get critical at~$z_s<10$. The solid and dashed curves represent the tangential and radial critical curves. The four different sets of curves are corresponding to four different source redshifts. At~$z_s=1.47$, the first HU gets critical, i.e.,~$z_{\rm hu,1}=1.47$. The second HU does not even get critical for a~$z_s=20$, and at~$z_s\geq10$, the critical curves effectively freeze.}
    \label{fig:plot_2enfw_g10}
\end{figure}

\begin{figure*}
    \centering
    \includegraphics[height=10.5cm, width=16cm]{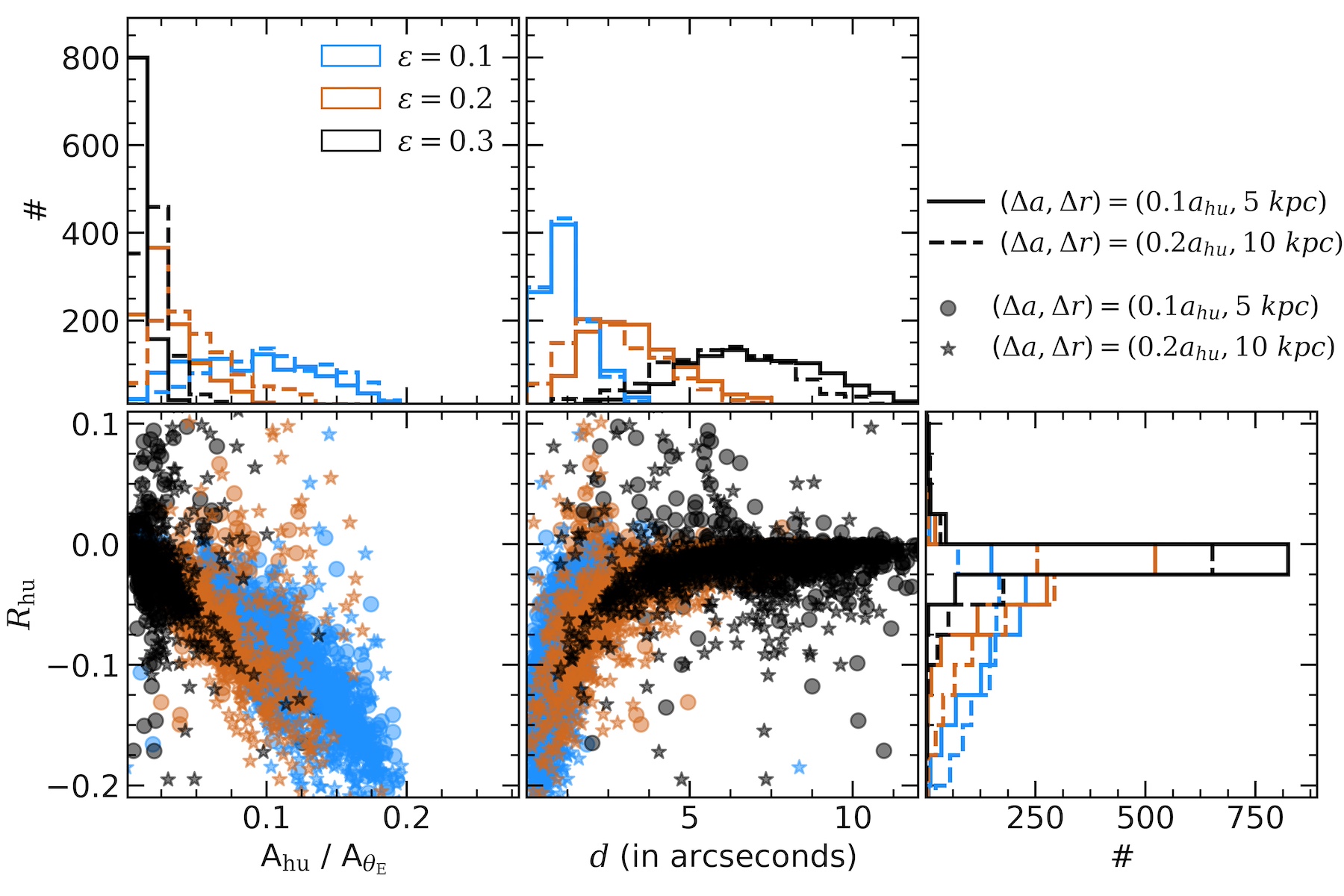}
    \caption{$R_{\rm hu, 1}$ in two-component eNFW lens. The left panel represents the~$R_{\rm hu, 1}$ as a function of the fraction of the area covered by the quadrilateral formed by four images with respect to the area covered by the Einstein ring~($A_{\rm hu}/A_{\theta_E}$). The middle panel shows the~$R_{\rm hu, 1}$ as a function of distance~($d$) of central maxima image from the lens centre. The histograms show different visualisations of the same information presented in scatter plots. In each panel, the blue, brown, and black points/histograms correspond to ellipticity~($\epsilon$) values of~0.3, 0.4, and 0.5, respectively. For each $\epsilon$-value, we chose two different sets of values for~$(\Delta a, \Delta r)$ to simulate lens systems, where~$\Delta a$ is the variation in distance ratio, and~$\Delta r$ is the radius of the circle from which the source position was drawn.}
    \label{fig:plot_2enfw_Rhu}
\end{figure*}

\section{Two-Component \lowercase{e}NFW Lens}
\label{sec:two_enfw}
Due to the reflection symmetry in a one-component eNFW lens, both HU points get critical at the same source redshift as we see in Figure~\ref{fig:hu_image} (for example image formations, see Figure~\ref{fig:example_1eNFW}). However, once we introduce external effects or a second lens component, this symmetry breaks, which in turn leads to both HU points getting critical at different source redshifts. In the presence of a second lens component, we may observe additional HUs~\citep[or other point singularities; see][]{2020MNRAS.492.3294M}. A two-component lens model is more realistic than a one-component lens model in the sense that even in the simplest cases of actual galaxy cluster lenses, there are always additional lens components (either in the form of substructures or in the form of external effects) which will break the reflection symmetry and make both HUs (of the primary lens component) get critical at different source redshifts. In this section, we study the effect of the presence of a second cluster-scale lens component on~$z_{\rm hu}$, magnification relation~($R_{\rm hu}$), and image formation near HUs of the primary lens component. We place a secondary cluster-scale eNFW lens component with~$(M_{\rm vir}, c_{\rm vir}, \epsilon)=(10^{14}~{\rm M_\odot}, 7, 0.3)$ randomly within an annulus of the inner radius of~$10''$ and width of~$20''$ centred on the primary lens. To conserve the total mass, we decrease the mass of the primary component to~$9\times10^{14}~{\rm M_\odot}$ and other lens parameters are the same as Section~\ref{sec:one_enfw}. This leads to an effective Einstein angle of~$\simeq27.5''$ at source redshift one. We limit the inner radius to~$10''$, keeping in mind that if we place a second cluster-scale lens component very close to the centre, it will significantly distort the overall caustic structure, and we may not be able to segregate HUs due to the primary lens component. On the other hand, the upper limit of~$30''$ makes sure that the effects due to the second component are not negligible. In this section, we use~$z_{\rm hu,1}$ and~$z_{\rm hu,2}$ for the primary and secondary HUs of the primary lens, respectively. By primary/secondary HU, we mean the primary lens component HU that gets critical first/second. The corresponding magnification relations are denoted by~$R_{\rm hu, 1}$ and~$R_{\rm hu,2}$, respectively.

\subsection{Effects on~$z_{\rm hu}$}
\label{ssec:2enfw_zhu}
The effects of the presence of a second cluster-scale lens component on~$z_{\rm hu}$ are shown in Figure~\ref{fig:plot_2enfw_zhu}. The x-axis represents the ratio of~$z_{\rm hu}$ value formed in the presence of the second lens component and~$z_{\rm hu, eNFW}$ value in single component lens case for a given~$\epsilon$ value, and the y-axis represents the distance~($d_{12}$) between the two lens components. From Figure~\ref{fig:plot_2enfw_zhu}, for~$\epsilon=0.1$, the presence of a second lens component does not significantly affect the~$z_{\rm hu}$ value as most of the~$z_{\rm hu}/z_{\rm hu, eNFW}$ values lie very close to one for both~$z_{\rm hu,1}$ and~$z_{\rm hu,2}$. That said, we do see a tail in the histogram plot at~$z_{\rm hu}/z_{\rm hu, eNFW}>1$. This behaviour can be understood from the fact that for~$\epsilon=0.1$, in one component lens, the HU gets critical for very small source redshift~($z_s=0.48$), and these redshifts the effect of the second component (which is at~$d_{12}\geq10''$) is not significant in the caustic structure, and they are still mostly the same as what we had in a single-component lens.

For~$\epsilon=0.2$ case, we see that the primary HU,~$z_{\rm hu, 1}$, gets critical at~$\lesssim z_{\rm hu, eNFW}$ (with peak at~$0.9 \times z_{\rm hu, eNFW}$) for~95\% cases and only~$\lesssim5\%$ of the systems get critical at~$>z_{\rm hu, eNFW}$. For~$z_{\rm hu, 2}$, nearly~60\% systems get critical at~$\lesssim z_{\rm hu, eNFW}$ and nearly~40\% of systems get critical at~$>z_{\rm hu, eNFW}$. A more significant effects occur for~$\epsilon=0.3$ where for~$z_{\rm hu, 1}$ nearly~85\% of the systems get critical at~$\lesssim z_{\rm hu, eNFW}$ (with 70\% systems getting critical at~$\simeq 0.5 \times z_{\rm hu, eNFW}$) and~15\% of systems get critical at~$>z_{\rm hu, eNFW}$. In addition, for~$\epsilon=0.3$, the fraction of~$z_{\rm hu,2}$ which get critical at~$\lesssim z_{\rm hu, eNFW}$ decreases to~40\%, and the fraction for~$>z_{\rm hu, eNFW}$ increases to~60\%.

From Figure~\ref{fig:plot_2enfw_zhu}, we see that adding a second lens component primarily decrease the~$z_{\rm hu, 1}$ for all values of~$d_{12}$. On the other hand, the effect of the second component on~$z_{\rm hu,2}$ is more complex. For~$\epsilon=0.3$, we see a large fraction of~$z_{\rm hu,2}$ which do not get critical up to~$z_{\rm hu,eNFW}$, nearly for all values of~$d_{12}$. This behaviour leads to the peak in the corresponding~$z_{\rm hu,2}$ histogram at~$z_{\rm hu,2}/z_{\rm hu,eNFW} \sim 4$. This peak marks the systems in which the~$z_{\rm hu,2}$ get critical at~$z_s\geq10$, and we argue that most of these systems will never get critical. The reason is very simple: at~$z_s\geq10$, the distance ratio is more or less constant, which implies that the caustic structure will also not evolve (or evolve extremely slowly). Hence, we have a very small chance of~$z_{\rm hu,2}$ getting critical at~$z_s\geq10$ if it did not get critical so far. As an example, in Figure~\ref{fig:plot_2enfw_g10}, we show critical curves for one system where~$z_{\rm hu,2}$ does not get critical even at~$z_s=10$. In Figure~\ref{fig:plot_2enfw_g10}, the primary HU gets critical at~$z_{\rm hu,1}=1.47$, and even at this redshift, we can get HU image formation near secondary HU (+ve y-axis), as can be implied from the close proximity and shape of radial and tangential critical curves. On the other hand,~$z_{\rm hu,2}$ does not get critical even at a source redshift of twenty, and the critical curves remain the same at~$z_s\gtrsim10$. Since, in such systems, HU does not get critical (or gets critical at very high redshifts), they can have large cross-sections for image formation near HU as the caustic structure nearly freezes. The same behaviour is also seen in actual cluster lenses, which is further discussed in Section~\ref{sec:actual_lens}. Hence, the presence of a second component mainly decreases the HU image formation cross-section near~$z_{\rm hu,1}$ but at the same time, a large fraction of systems can have large HU image formation cross-sections around~$z_{\rm hu,2}$ especially for~$\epsilon\geq0.3$.

\begin{figure*}
    \centering
    \includegraphics[height=6.8cm, width=18cm]{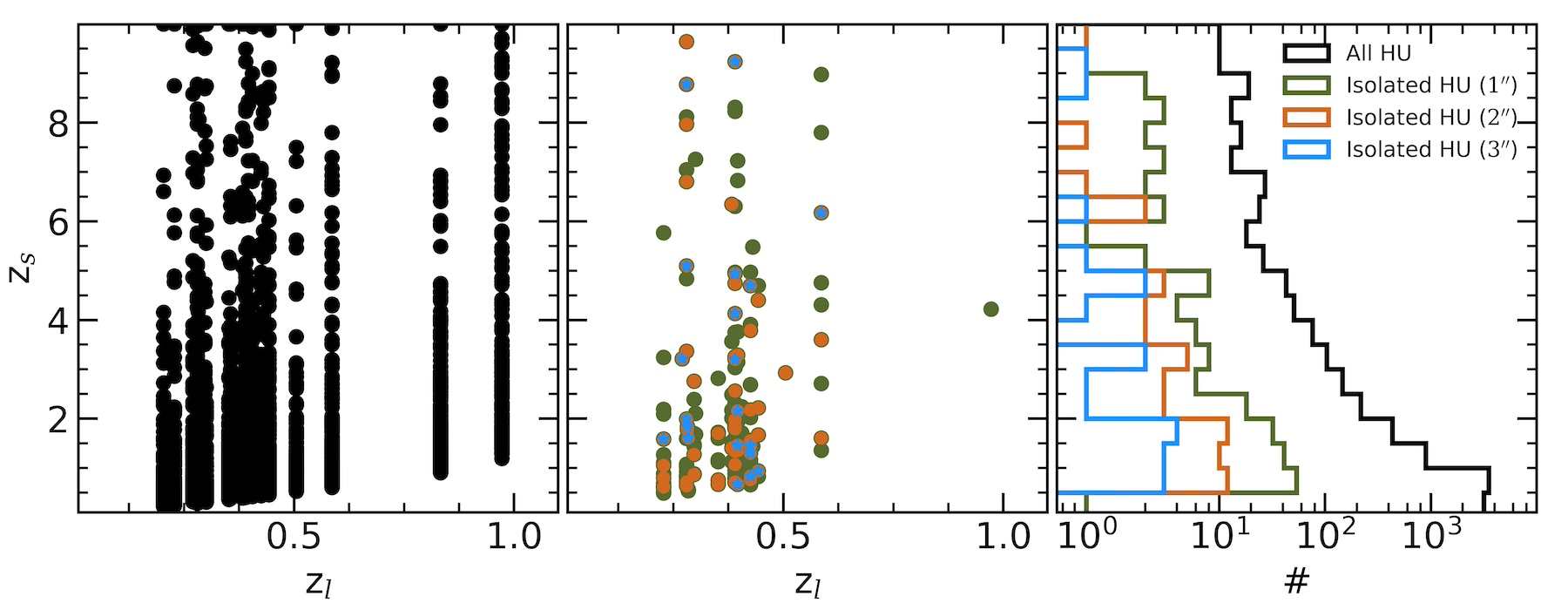}
    \caption{Redshift distribution of HU singularities in a sample of real cluster lenses (shown in Table~\ref{tab:clusters}). In the left and middle panels, the x- and y-axis denote the cluster redshift and HU redshift. The left panel represents all of the HU points that get critical below a source redshift of ten. In the middle panel, green, brown, and blue points show the isolated HUs such that within~$1''$,~$2''$, and~$3''$, respectively, no other HU point lies. The corresponding histograms are shown in the right panel using the same color.}
    \label{fig:plot_zhu_real}
\end{figure*}

\subsection{Effects on~$R_{\rm hu,1}$}
\label{ssec:2enfw_Rhu}
Effects of the presence of a second lens component on the~$R_{\rm hu,1}$ corresponding to~$z_{\rm hu, 1}$ are shown in Figure~\ref{fig:plot_2enfw_Rhu}. Here, we again trace the cusp with two different sets of~$(\Delta a, \Delta r)$ to determine the source redshift~($z_s$) and its position around the cusp point. We can observe that the overall trend of low ellipticity, giving negative~$R_{\rm hu,1}$ values and by increasing the ellipticity shifts to zero value,~i.e.,~$R_{\rm hu,1}\simeq0$, holds true. However, the presence of a second lens component and the corresponding decrease in the~$z_{\rm hu,1}$ introduce scatter in the~$R_{\rm hu,1}$ values compared to the single eNFW case. In addition, for~$\epsilon=0.2$ case, we observe an extended tail (compared to single eNFW lens) in~$R_{\rm hu,1}$ on negative values, resulting from the decrease in~$z_{\rm hu,1}$ values. For~$\epsilon=0.3$ case, we still see that nearly~$\sim85\%$ systems give~$|R_{\rm hu,1}| < 0.05$ for both cases of~$(\Delta a, \Delta r)$ but some cases lead to~$|R_{\rm hu, 1}|$ as large as 0.1. Due to the decrease in~$z_{\rm hu,1}$ value, the distance~($d$) of the central image from the lens centre also decreases and can have values in a broader range of~${\simeq}[2.5'', 12'']$ which also contributes to the scatter in~$R_{\rm hu,1}$ values (primarily shifting towards negative values).

The other reason that can introduce deviations in ~$R_{\rm hu,1}$ from a zero value is the additional caustics introduced by the second lens component. A hint of this can again be seen in Figure~\ref{fig:plot_2enfw_g10} where we can see the distortion in both radial and tangential critical curves and these distortions will also be reflected in the caustics. If the second component introduces additional caustics close to the HU point, for example, by giving rise to a new swallowtail on the existing caustics or by introducing additional secondary caustics, it will primarily affect the magnification values of one/two images since the second component is an order of magnitude smaller than the primary caustic and introduce deviations in the~$R_{\rm hu,1}$ value. Specifically, for Figure~\ref{fig:plot_2enfw_g10} case, the second component lies on the negative x-axis side and it will introduce maximum variation in the magnification of the saddle-point image forming closest to it. If it (de-)magnifies the saddle-point image further, the~$R_{\rm hu,1}$ will move towards (positive)~negative values.

\section{Actual cluster lenses}
\label{sec:actual_lens}
An actual galaxy cluster lens, in principle, has non-zero ellipticity and contains a large number of galaxy-scale substructures (which also have non-zero ellipticities). Hence, we can expect that every cluster-scale lens has the capacity to lead HU image formation. That said, the spatial and redshift distribution of these HUs need to be determined since only then can we estimate the cross-section of the HU image formation in a given sample of cluster lenses. In this section, we study the~$z_{\rm hu}$ distribution in actual cluster lenses and search for cluster lenses with large HU image formation cross-sections in twenty of the (randomly selected) RELICS clusters along with RXJ0437 and Abell~1703. The complete list of clusters used here is shown in Table~\ref{tab:clusters}.

\subsection{$z_{\rm hu}$ distribution}
\label{ssec:actual_zhu}
In cluster-scale lenses, the primary mass contribution comes from the dark matter~(DM) halo and the cluster galaxies mainly provide additional lensing perturbations to the overall cluster lensing~(unless a large DM halo is associated with one of these galaxies). Every cluster galaxy will, at least, lead to one pair of HU singularities and the corresponding~$z_{\rm hu}$ values will depend on the ellipticity and the position of the galaxy within the cluster~(since the position determines the overall lensing effect of the cluster on the cluster galaxy). Similarly, the DM halo will also lead to at least a pair of HUs.

The distribution of all HUs in all of the clusters (mentioned in Table~\ref{tab:clusters}) is shown in Figure~\ref{fig:plot_zhu_real}. We can see that the total number of HU in the current sample of clusters is more than~$5\times10^3$ (black points in the left panel and black histogram in the right panel). However, most of these HUs get critical at very small redshifts (close to the corresponding cluster redshift), implying that they mainly originated from the cluster galaxies. The smaller redshift also implies that the corresponding cross-section to give rise to an HU image formation is very small since these HUs get critical for smaller redshifts, and the position of these HUs will be very close to the centre of cluster galaxies. Even at large redshifts, if the HUs are originating from the cluster galaxies, the corresponding cross-section will be small since the source needs to lie in a very small region around HU to give rise to an HU image formation. Otherwise, the image formation will look more like the typical quad image formation. One can try to remove the HU contribution from cluster galaxies by masking the regions at each galaxy; however, since we are only working with mass models, we cannot perform this. Instead, we use a different method. In addition to forming close to the cluster galaxies, cluster galaxy HU pairs will also lie close to each other, so we can remove such pairs by looking for isolated HUs within a certain radius. The green, brown, and blue points in the middle panel show the isolated HUs such that within~$1''$,~$2''$, and~$3''$ radius, no other HU point lies. We can see that the number of isolated HUs decreases significantly, and only~${\sim}5\%$,~${\sim}1\%$, and~${\lesssim}0.5\%$ HU points have no other HUs within~$1''$,~$2''$, and~$3''$ radius, respectively. With this, the average number of isolated HUs within~$3''$ radius, which may have large image formation cross-sections, in each cluster is~$\sim1$ as the total number of such HUs is~$\sim20$. And, as expected, this number continuously drops as we go towards higher source redshifts.

\begin{figure*}
    \centering
    \includegraphics[height=10.5cm, width=16cm]{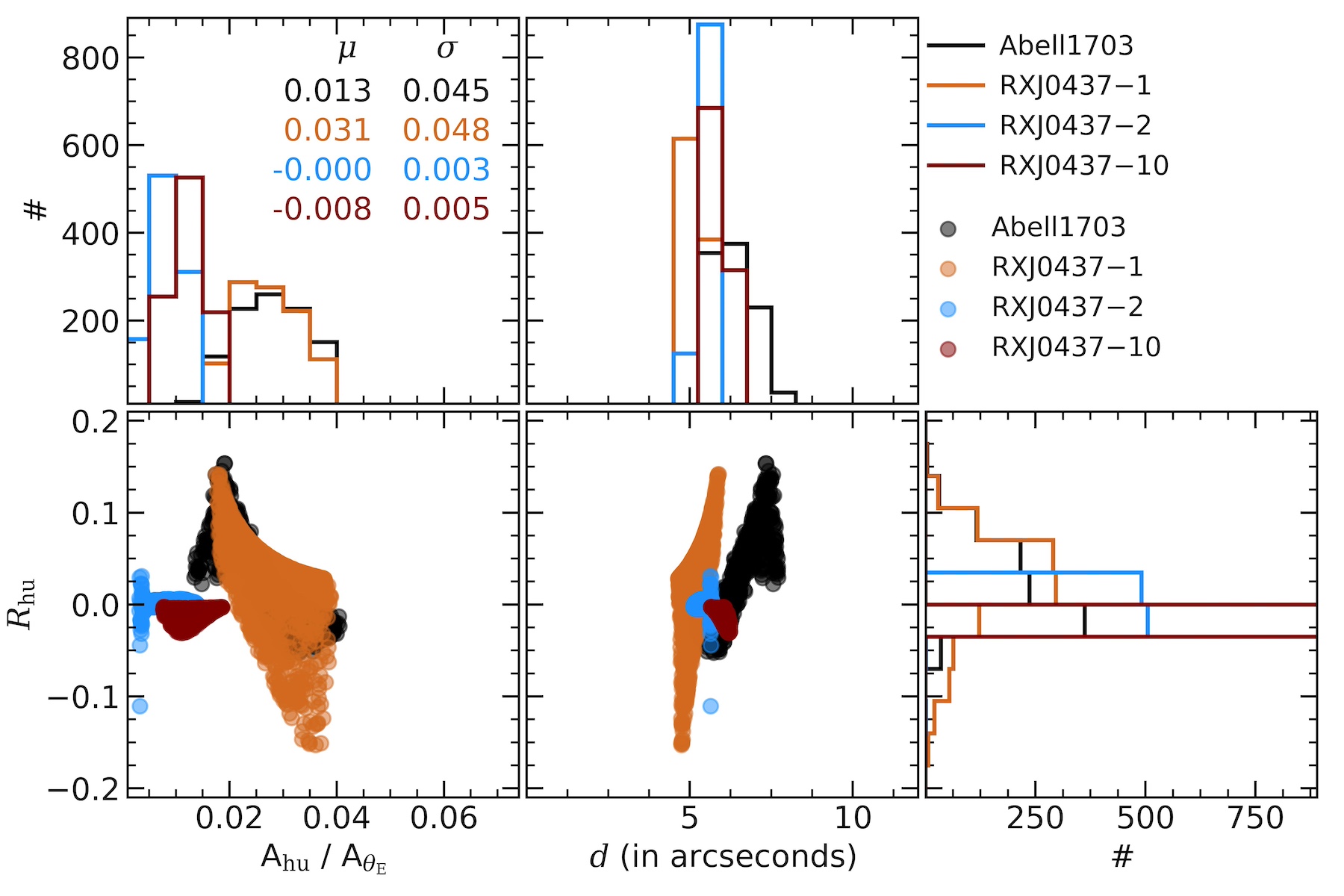}
    \caption{$R_{\rm hu}$ distribution near observed HU image formations in Abell~1703 and RXJ0437 clusters. In Abell~1703, we only have one system leading to HU image formations, whereas RXJ0437 led to three HU image formations. Various panel axes are similar to Figure~\ref{fig:plot_1enfw_hu} and~\ref{fig:plot_2enfw_Rhu}. The median and one sigma scatter in~$R_{\rm hu}$ values for each system are shown in the top left panel with a colour scheme the same as the scatter plots.}
    \label{fig:plot_Rhu_real}
\end{figure*}

The above analysis does hint at one reason why it is relatively rare to observe image formation near HU singularities, even though their overall number can be very large in cluster lenses. However, it is important to point out that isolated HUs, for example, let us say within~$3''$ radius may also remove HUs that are generated from the cluster DM halo. This can happen if we have an HU singularity due to the DM halo, but a small cluster galaxy is sitting close to it. In such case, this HU will not be part of the sample (since it is not isolated) but it can lead to HU image formation and the cluster galaxy will primarily introduce perturbation in the image formation and its properties. Hence, the above segregation method is expected to give us a lower limit on the number of HUs which can have large image formation cross-sections.

\subsection{Clusters with large HU image cross-section}
\label{ssec:actual_cs}
Similar to the double component lens discussed in Section~\ref{sec:two_enfw}, real cluster lenses can also have large cross-sections for image formation near HUs. Such clusters can easily be identified if the corresponding (radial and tangential) critical curves show similar behaviour as the critical curves close to the HU singularity and are frozen against changes in distance ratio. Out of twenty RELICS clusters shown in Table~\ref{tab:clusters}, at least seven clusters have large cross-sections for HU image formations. The critical curves for these seven clusters are shown in Figure~\ref{fig:large_cs} and the last panel shows the RXJ0437 cluster lens, which also shows large cross-sections for HU image formations. In each panel of Figure~\ref{fig:large_cs}, we can see that the critical curves, as we go from~$z_s=5$ to~$z_s=10$, remain nearly frozen and the geometry of critical curves is similar to HU singularity close to the BCG. If we assume that only these seven clusters have large cross-sections for HU image formations in the whole RELICS sample, we can roughly imply from this that one to two in ten clusters will have a large cross-section for HU image formation. Although no confirmed HU image formations are observed in the RELICS sample (except one/two candidates mentioned in~\citealt{2023MNRAS.526.3902M}), such clusters can be excellent targets for deep imaging with JWST to search for HU image formation at large redshifts. In addition, as discussed in~\citet{2023MNRAS.522.1091L}, some of the HU image formations are only detected with Multi-Unit Spectroscopic Explorer~(MUSE) on the Very Large Telescope (VLT) which is very sensitive to the sources with emission features, which may be extremely faint in broadband imaging, another potential reason for the small number of observed image formations near HU.

Here, it is (again) important to highlight that, the above inference made in this section regarding the distribution of HU singularities and that some of these clusters have large cross-sections for HU image formation also depends on the underlying lens model since different lens models can generate different critical curves and caustics with the degree of variation depending on the nature of the lens modelling method~\citep[e.g.][]{2017MNRAS.472.3177M}. However, since the underlying models used in our current work are all parametric (except for Abell~1703 where we use~\textsc{ltm}), the above inferences are credible. 

\begin{figure*}
    \centering
    \includegraphics[height=8cm, width=16cm]{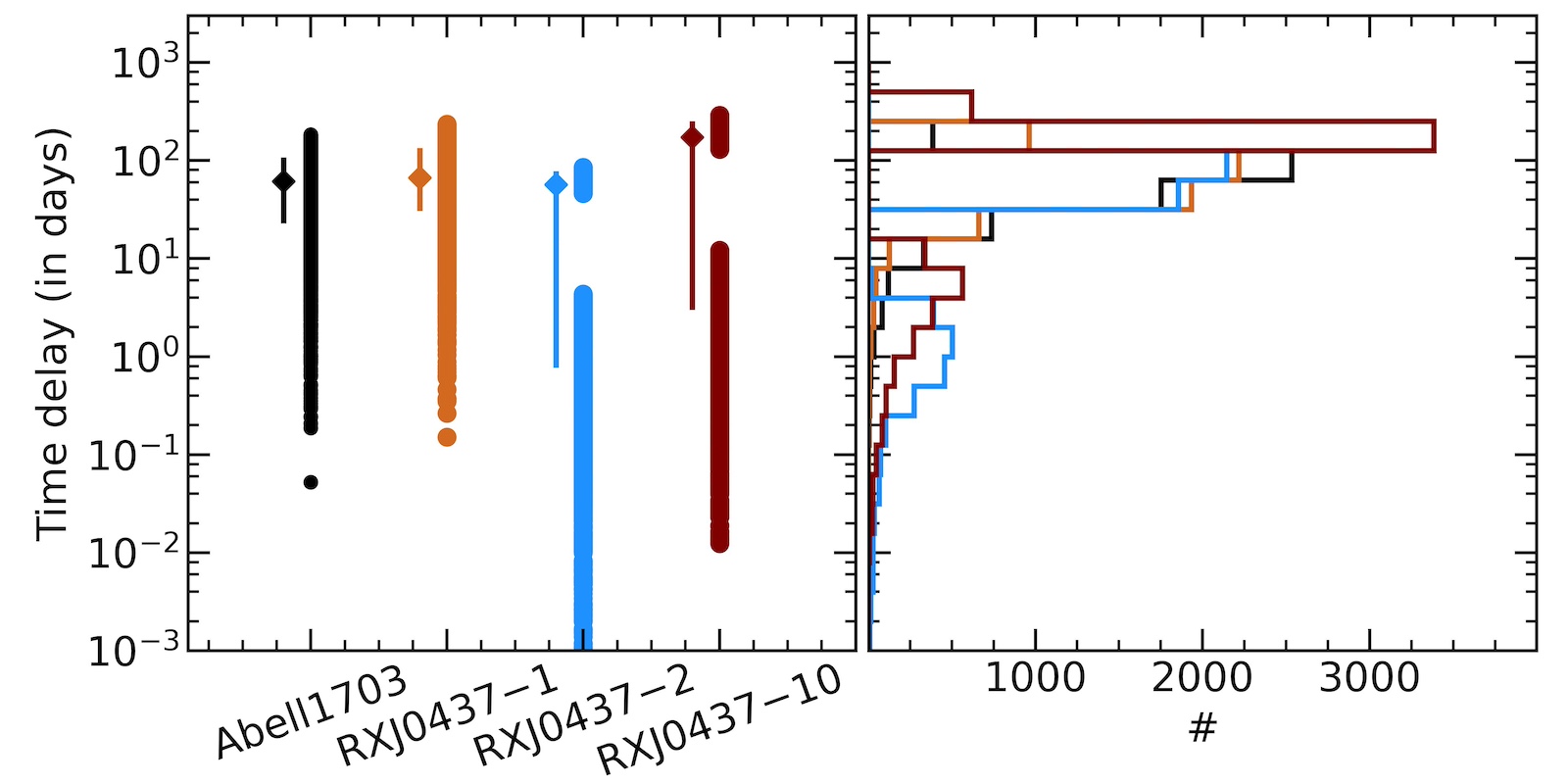}
    \caption{Time delay distribution near observed HU image formations in Abell~1703 and RXJ0437. The left represents the scatter plot of the time delay values between HU images. For each HU image formation, we plot six time delay pairs. The x-axis represents the name of the cluster and HU image system. For each HU image system, the diamond point with the error bar shows the median value and the 16th and 84th percentile region around it. The right panel shows the histogram corresponding to the points shown in the left panel.}
    \label{fig:plot_TD_real}
\end{figure*}

\section{Observed HU image formations}
\label{sec:obs_hu}
So far, four HU image formations have been discussed in the literature in detail in two different galaxy clusters, namely, Abell~1703~\citep{2008A&A...489...23L} and RX~J0437.1+0043~\citep[RXJ0437;][]{2023MNRAS.522.1091L}. Other possible candidates are presented in~\citet{2023MNRAS.526.3902M} and~\citet{2024arXiv240411659E}. In this section, we restrict our attention to the four image formations that have been observed. Image formation observed in Abell~1703 and for system-1 in RXJ0437 are ideal HU image formation showing ring-like image formations, whereas the other two image formations in RXJ0437 corresponding to system-2 and system-10, are variations of ideal HU image formation when the source lies close to both radial and tangential caustics simultaneously~(see Figure~12 in~\citealt{2023MNRAS.522.1091L}). In this section, we study the~$R_{\rm hu}$ and time-delay distribution in these image formations. To do so, for each of these image formations, we take an ellipse situated on one of the lensed images, which is part of HU image formation and determine the corresponding source position, and using this source position, we determine the rest of the image positions~(similar to the analysis done in Section~7 of~\citealt{2023MNRAS.526.3902M}). This way, for each observed HU image formation, we simulate~$10^3$ lens systems. For Abell~1703, we use the lens model constructed using the \textsc{light-trace-mass}~(\textsc{ltm}) presented in~\citet{2010MNRAS.408.1916Z} and for RXJ0437, we use the parametric \textsc{Lenstool} lens model from~\citet{2023MNRAS.522.1091L}.

\subsection{$R_{\rm hu}$ distribution}
\label{ssec:obs_rhu}
The~$R_{\rm hu}$ distribution for simulated lens systems around the observed HU image formations are shown in Figure~\ref{fig:plot_Rhu_real}. We can see that the area covered by the HU quadrilateral is~$\lesssim4\%$ of the area corresponding to the effective Einstein angle,~$A_{\theta_E}$, and the distance of the central image from the lens centre is~$\gtrsim5''$ for all four of these image formations. System-2 and system-10 in RXJ0437 cover very small area values and also have~$R_{\rm hu}\simeq0$. The median~$R_{\rm hu}$ values and $1-\sigma$ scatter (covered by 16th and 84th percentile values) around it are shown in the top-left panel with the colour scheme the same as the scatter plots. Both of these behaviours (small~$A_{\theta_E}$ and~$R_{\rm hu}$) can be understood from the fact that these image formations have pairs of images lying close to each other near the critical curve, implying that each pair will have similar magnification values leading to~$R_{\rm hu}$ very close to zero, and since we are looking at the area covered by the quadrilateral made of these four images, the area will also be smaller. System-1 in RXJ0437 shows the largest scatter in~$R_{\rm hu}$ with $(\mu,\sigma)=(0.031, 0.048)$, which could be stemming from the presence of cluster galaxies close to the central maxima image~(see Figure~6 in~\citealt{2023MNRAS.522.1091L}). The same reason~(i.e., the presence of cluster galaxies) can also be given for Abell~1703, where again we have cluster galaxies close to the image formation~(see Figure~9 in~\citealt{2023MNRAS.526.3902M}).

\subsection{Time delay distribution}
\label{ssec:obs_td}
In general, when multiple lensed images lie close to each other, they have large magnification and small relative time delays compared to images lying far from each\footnote{Lenses with certain symmetries can have smaller time delays (and large magnifications) even when lensed images lie very far from each other.}. The same is also expected to be true for HU image formation, where four images lie close to each other~\citep{2022MNRAS.515.4151M}. Since HU image formation contains four images, we have six time delay values. Hence, for~$10^3$ simulated systems, we have~$6\times10^3$ time delay values. The time delay distributions around the four HU image formations discussed above are shown in Figure~\ref{fig:plot_TD_real}. Except for RXJ0437-10, for the other three image formations, the median time delay value is~$\simeq 80$~days with the 84th percentile remaining~$\lesssim200$~days. For RXJ0437-10, the median is~$\simeq200$~days and the 84th percentile value is~$\lesssim300$~days. In the histogram plot, for RXJ0437-2 and RXJ0437-10, we see a secondary~(less significant) peak around~$1-10$~days, which is not seen in the other two systems. This is due to the fact that in RXJ0437-2 and RXJ0437-10 pair of images lie very close critical curves (and to each other), and the time delay between pair of images near to each other remains very small.

Four images and small time delays imply that any intrinsic variation in the source plane will be observed four times within less than a year, opening a new avenue to perform time-delay studies. More so with HUs with active star formation, such as in RXJ0437-1 and RXJ0437-2~\citep{2023MNRAS.522.1091L}. Since HUs have images very close to the critical curves, another source of variability in HUs can come from caustic-crossing events~\citep[e.g.,][]{2018NatAs...2..334K} where a background star crosses a micro-caustic, formed due to microlenses in the intra-cluster medium. Although, such variability will not be seen in all images. It is noteworthy that the shorter time delay estimates also require a review of the cadence for observations of any transients associated with the lensed sources.

\section{Conclusions}
\label{sec:conclusions}
In this work, we have studied the properties of image formation near HU singularities in simple simulated and real galaxy cluster-scale lenses, primarily variation in corresponding magnification relation~($R_{\rm hu}$) as a function of the area covered by the image formation and the distance of central maxima image from the lens centre. We find that for~$\epsilon=0.3$ in an eNFW lens, the area covered by the four images quadrilateral is~$\lesssim5\%$ of the area covered by the effective Einstein radius for~$(\Delta a, \Delta r)\leq (0.2 a_{\rm hu}, 10{\rm kpc})$ and the distance of central maxima always remains~$\gtrsim5''$ from the lens centre while~$R_{\rm hu}$ always remains very close to zero~(i.e.,~$R_{\rm hu}\approx0$). In addition, with the above specification, it is also straightforward to identify the HU image formation visually. Hence, the above~$(\Delta a, \Delta r)$ can be used to estimate the lower limit on the HU cross-section for future surveys. On the other hand, having a smaller ellipticity can deviate the~$R_{\rm hu}$ considerably from zero as the central maxima gets de-magnified and can be un-observable in real lenses due to the presence of BCG light. Adding a second cluster-scale lens component primarily brings down the primary HU critical redshift~($z_{\rm hu, 1}$) corresponding to the primary lens component, leading to a decrease in the HU image formation cross-section. However, at the same time for~$\epsilon\geq0.3$, often, the presence of a second component does not let~$z_{\rm hu,2}$ get critical which in turn leads to large cross-sections for image formations near HU, and this effect is not limited to the ideal lenses as eight out of twenty-two clusters used in our current work show similar behaviour. This implies that a large population of clusters (one to two in every ten clusters based on the above numbers) can have large cross-sections for HU image formations, potentially making such image formations less rare in future surveys.

In four actual observed HU image formations, which are studied in detail, we find that the central maxima lie at a distance of~$\gtrsim5''$ from the lens centre similar to what we observe in ideal cases for~$\epsilon=0.3$. In two of the HU systems, we have median~$|R_{\rm hu}|<0.001$ with 1-$\sigma$ scatter~$\leq0.005$ and the other two systems have median~$|R_{\rm hu}|<0.05$ and 1-$\sigma$ scatter of~$\leq0.05$. The higher scatter in the later systems is because of the presence of cluster galaxies near HU images. The median value of relative time delay between HU image formations in all of these systems is~$<200$~days, with some pairs (close to the critical curve) having values less than a day. The time delay analysis for these four known HU systems suggests that the time delay for transients associated with such source galaxies will be an order of magnitude smaller than that for generic five image systems in cluster lensing. For some pairs, the time delay may be as small as a few weeks. Exploring this observationally will require a different approach as the shorter time delays require a higher cadence and possibly monitoring from space or using telescopes around the globe.

We expect that many more HU systems will be discovered in the coming years and this will allow us to analyze the statistics of these systems more reliably. Our earlier studies have shown that different approaches to making lens maps lead to very different estimates for the number of HU systems that are likely to be discovered~\citep{2021MNRAS.503.2097M, 2022MNRAS.515.4151M}. Thus, discoveries of HU systems will also tell us about which method for modelling yields predictions that match with observations. To efficiently identify such HU image formation in large surveys, one can rely on automated searches, which are already in use to find strong lensing systems~\citep[][]{2018MNRAS.473.3895L, 2019MNRAS.487.5263D, 2024A&A...681A..68E} and is part of our ongoing research.

\section{Data Availability}
The simulated data used in the current work can be easily generated with the methods discussed in the text. Lens models for Abell~1703 and RXJ0437 can be made available upon request to modellers.

\section*{Acknowledgements}
Authors thank David Lagattuta, Marceau Limousin, and Adi Zitrin for providing lens models for Abell~1703 and RXJ0437 galaxy clusters. AKM thanks Joseph Allingham, Guillaume Mahler, and Johan Richard for help with \textsc{Lenstool} software. Authors thank the anonymous referee for the helpful comments. AKM acknowledges support by grant 2020750 from the United States-Israel Bi-national Science Foundation (BSF) and grant 2109066 from the United States National Science Foundation (NSF) and the Ministry of Science $\&$ Technology, Israel. JSB thanks NCRA-TIFR for its hospitality, as this manuscript was completed during the sabbatical from IISER Mohali. 

This research has made use of NASA's Astrophysics Data System Bibliographic Services. This work is based on observations taken by the RELICS Treasury Program~(GO 14096) with the NASA/ESA HST, which is operated by the Association of Universities for Research in Astronomy, Inc., under NASA contract NAS5-26555.

This work utilises the following software packages:
\textsc{python}~(\url{https://www.python.org/}),
\textsc{astropy}~\citep{2018AJ....156..123A},
\textsc{matplotlib}~\citep{2007CSE.....9...90H},
\textsc{NumPy}~\citep{2020Natur.585..357H},
\textsc{SciPy}~\citep{2020NatMe..17..261V},
\textsc{shapely}~\citep{gillies_sean_2022_7428463},
\textsc{Lenstool}~\citep{1996ApJ...471..643K}.

\begin{table}
    \centering
    \caption{\justifying Galaxy clusters used in Section~\ref{ssec:actual_cs}. Columns~(1), (2) and~(3) represent the cluster name, cluster redshift, and corresponding lens models used in our current work, respectively. Column~(4) represents the cluster survey catalogue from where these lens models are taken from. Except for the last two cluster lenses, all other clusters are taken from RELICS.}
    \begin{tabular}{cccc}
    \hline
    Cluster name       & $z_l$    & lens model         & Catalogue      \\
    (1)                &  (2)     & (3)                & (4)            \\
    \hline    
    Abell 1758         & 0.280    & \textsc{glafic}    & RELICS         \\
    Abell 2163         & 0.203    & \textsc{glafic}    & RELICS         \\
    Abell 2537         & 0.297    & \textsc{glafic}    & RELICS         \\
    Abell 3192         & 0.425    & \textsc{glafic}    & RELICS         \\
    Abell 697          & 0.282    & \textsc{glafic}    & RELICS         \\
    Abell S295         & 0.300    & \textsc{glafic}    & RELICS         \\
    CLJ0152.7-1357     & 0.833    & \textsc{glafic}    & RELICS         \\
    MACS J0025.4-1222  & 0.586    & \textsc{glafic}    & RELICS         \\
    MACS J0035.4-2015  & 0.352    & \textsc{glafic}    & RELICS         \\
    MACS J0159.8-0849  & 0.405    & \textsc{glafic}    & RELICS         \\
    MACS J0257.1-2325  & 0.505    & \textsc{glafic}    & RELICS         \\
    MACS J0308.9+2645  & 0.356    & \textsc{glafic}    & RELICS         \\
    MACS J0417.5-1154  & 0.443    & \textsc{glafic}    & RELICS         \\
    MACS J0553.4-3342  & 0.430    & \textsc{glafic}    & RELICS         \\
    PLCK G171.9-40.7   & 0.270    & \textsc{glafic}    & RELICS         \\
    PLCK G287.0+32.9   & 0.390    & \textsc{glafic}    & RELICS         \\
    RXC J0032.1+1808   & 0.396    & \textsc{glafic}    & RELICS         \\
    RXC J0949.8+1707   & 0.383    & \textsc{glafic}    & RELICS         \\
    RXS J060313.4+4212S& 0.228    & \textsc{glafic}    & RELICS         \\
    SPT-CLJ0615-5746   & 0.972    & \textsc{glafic}    & RELICS         \\
    Abell 1703         & 0.282    & \textsc{ltm}       & --             \\
    RX J0437.1+0043    & 0.285    & \textsc{Lenstool}  & --             \\
    \hline
    \end{tabular}
    \label{tab:clusters}    
\end{table}  

\bibliographystyle{mnras}
\bibliography{reference}

\begin{figure*}
    \centering
    \includegraphics[height=5cm, width=5cm]{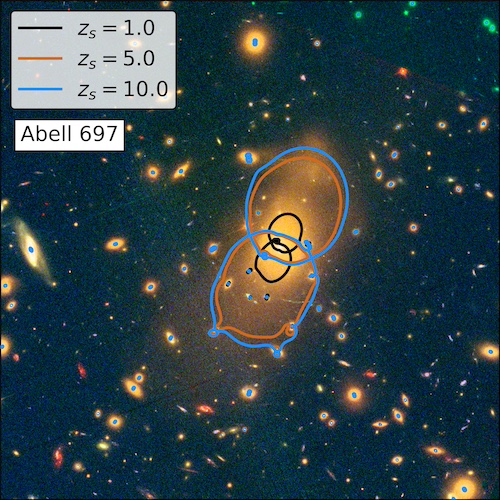}
    \includegraphics[height=5cm, width=5cm]{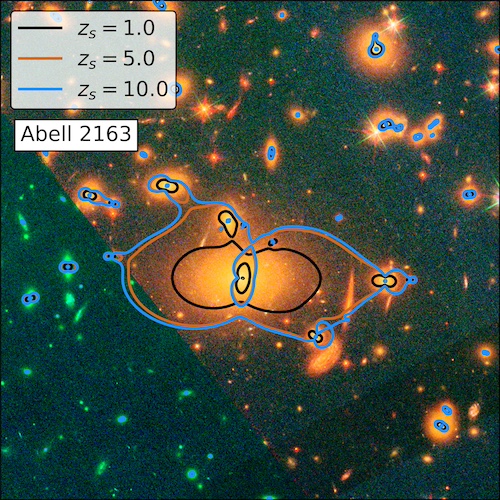}
    \includegraphics[height=5cm, width=5cm]{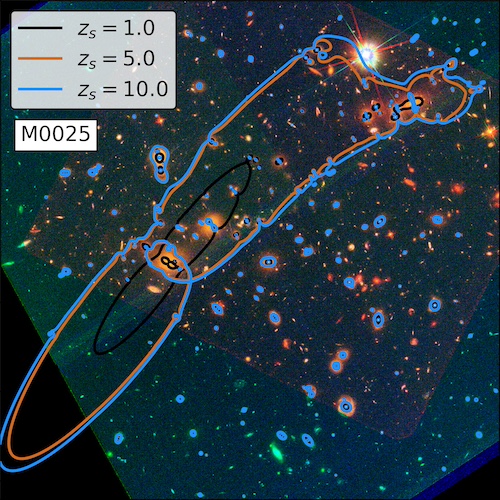}
    \includegraphics[height=5cm, width=5cm]{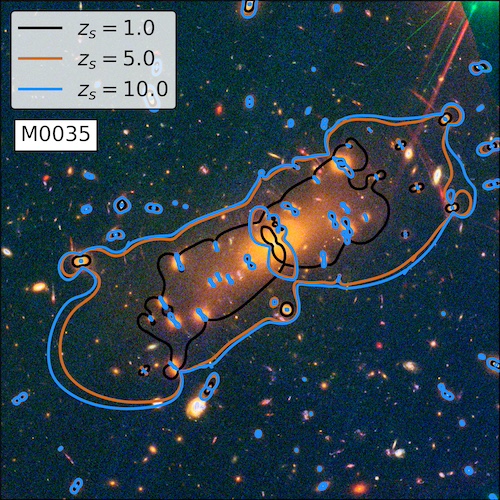}
    \includegraphics[height=5cm, width=5cm]{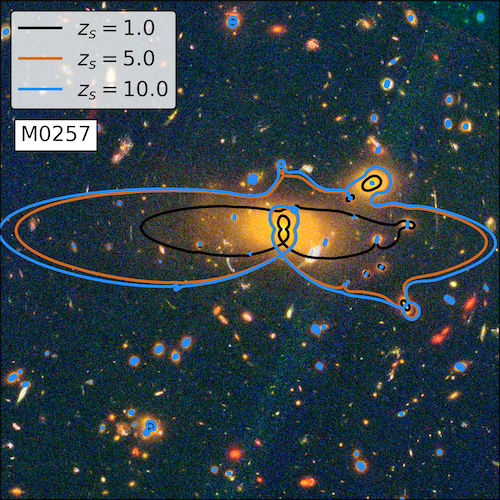}
    \includegraphics[height=5cm, width=5cm]{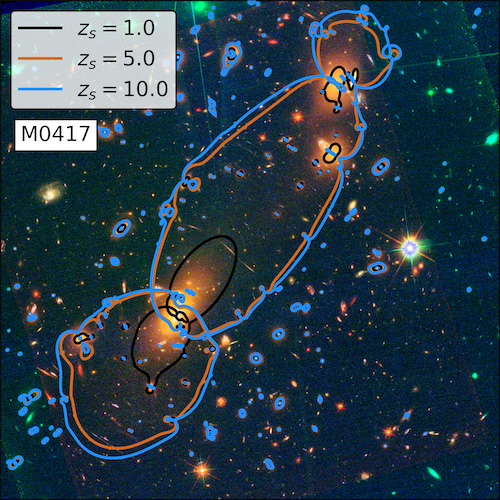}
    \includegraphics[height=5cm, width=5cm]{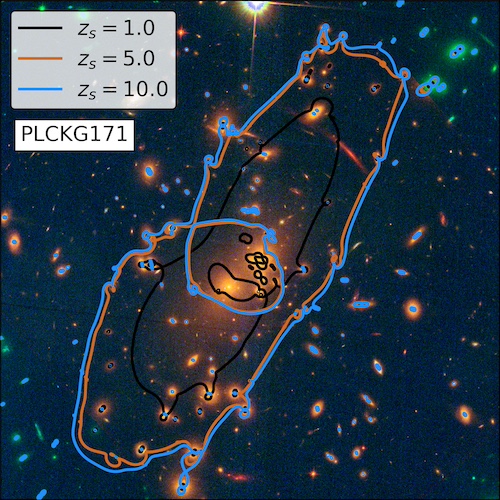}
    \includegraphics[height=5cm, width=5cm]{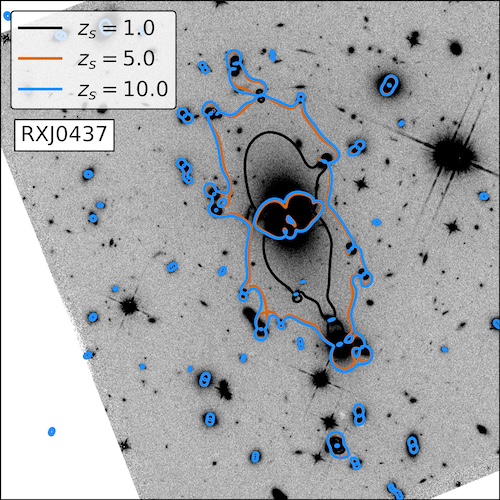}
    \caption{Actual clusters with large cross-section for HU image formation. Except for the last panel~(RXJ0437), all other cluster lenses are taken from the RELICS. In each panel, the black, brown, and blue curves represent the critical curves for a source redshift of~1,~5, and~10, respectively.}
    \label{fig:large_cs}
\end{figure*}

\begin{figure*}
    \centering
    \includegraphics[height=20cm, width=15cm]{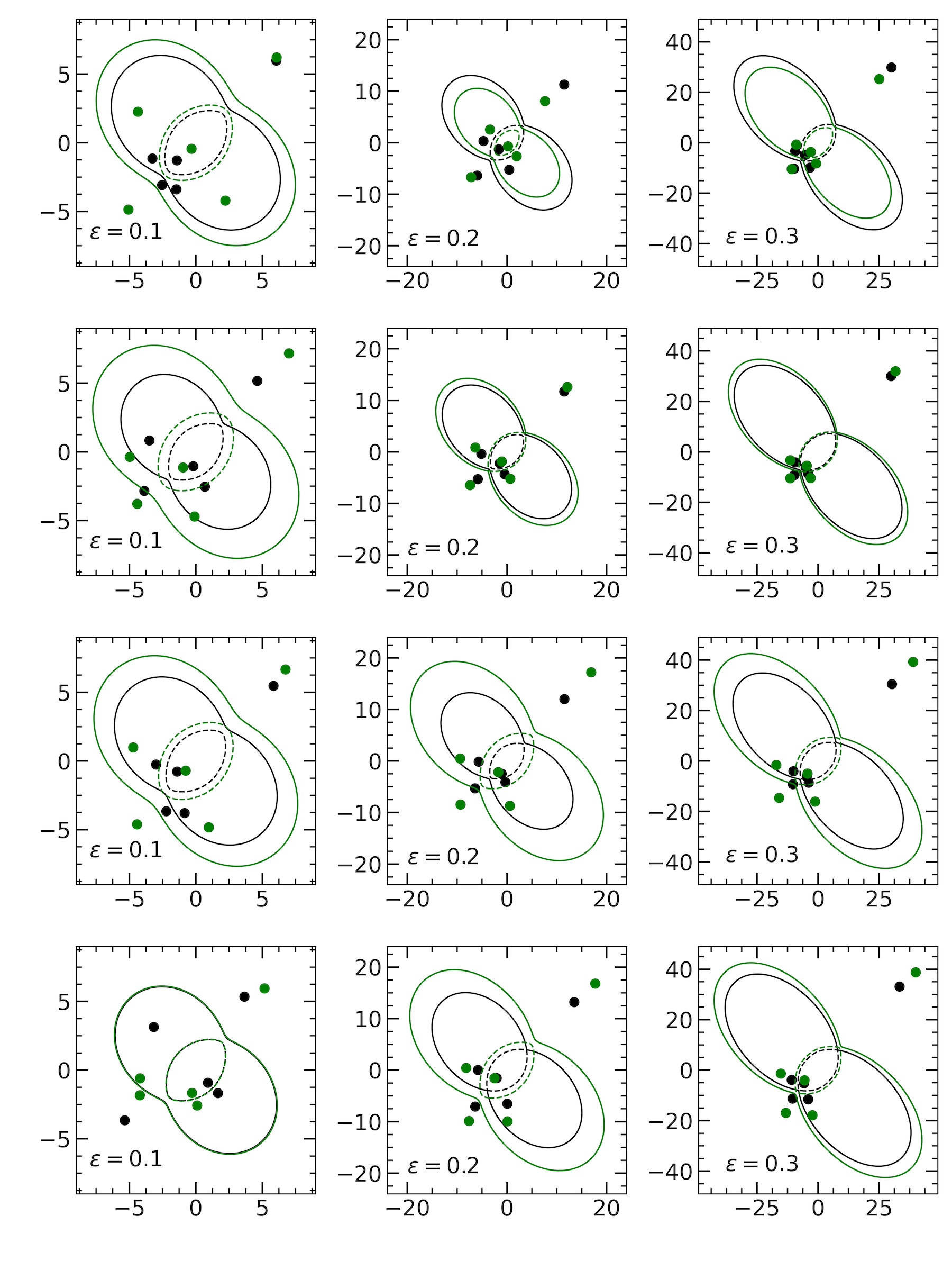}
    \caption{Example of HU image formation in one-component eNFW lens. The left, middle, and right columns represent the randomly chosen cases for~$\epsilon=0.1, 0.2$, and~0.3, respectively. In each panel, the black solid and dashed curves represent the tangential and radial critical curve for~$(\Delta a, \Delta r) = (0.1 a_{\rm hu}, 5~{\rm kpc})$. Green critical curves are corresponding to~$(\Delta a, \Delta r) = (0.1 a_{\rm hu}, 10~{\rm kpc})$. The corresponding image formations are shown in the same colour.}
    \label{fig:example_1eNFW}
\end{figure*}

\begin{figure*}
    \centering
    \includegraphics[height=20cm, width=15cm]{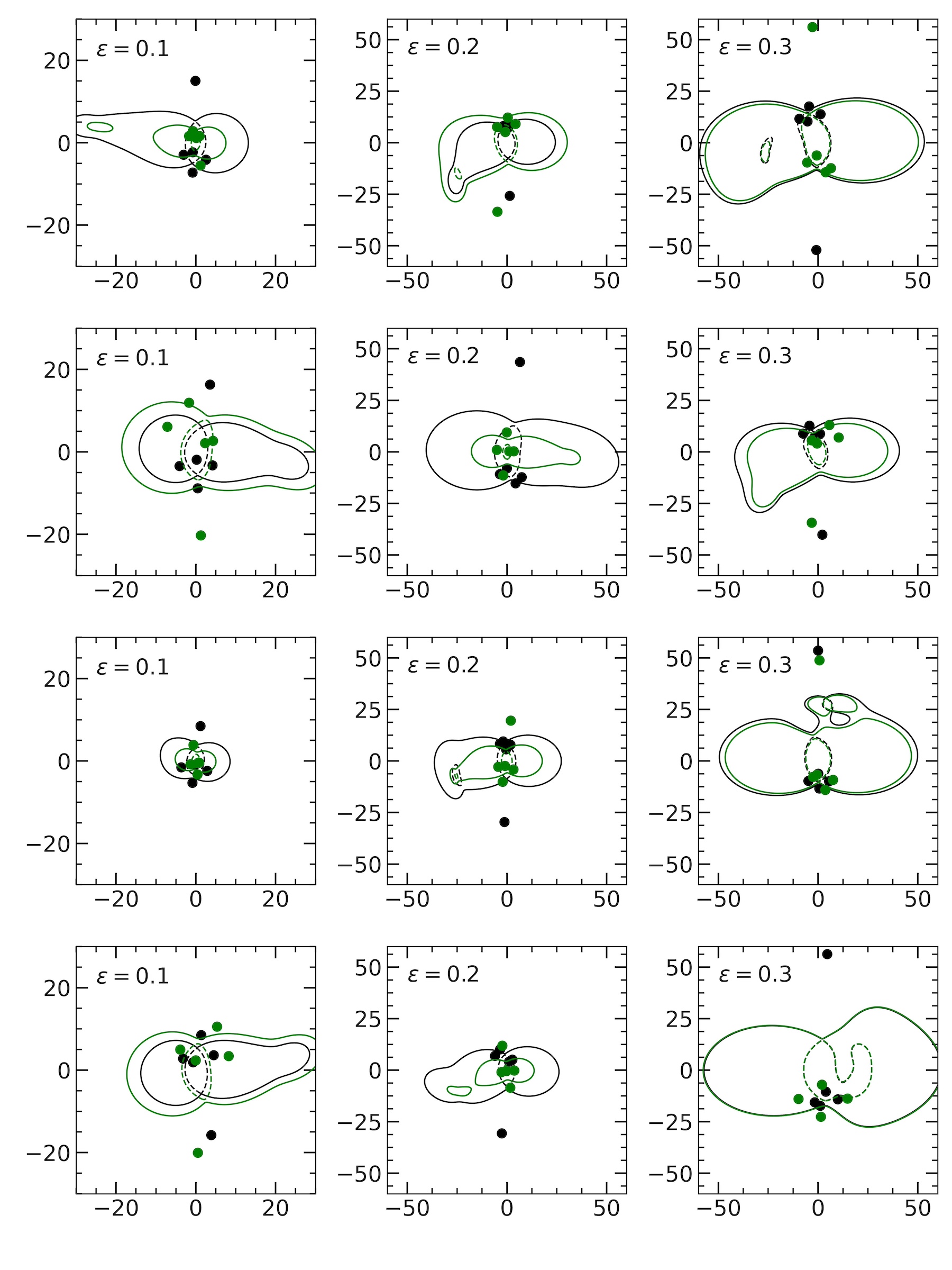}
    \caption{Example of HU image formation in two-component eNFW lens. The left, middle, and right columns represent the randomly chosen cases for~$\epsilon=0.1, 0.2$, and~0.3, respectively. In each panel, the black solid and dashed curves represent the tangential and radial critical curve for~$(\Delta a, \Delta r) = (0.1 a_{\rm hu}, 5~{\rm kpc})$. Green critical curves are corresponding to~$(\Delta a, \Delta r) = (0.1 a_{\rm hu}, 10~{\rm kpc})$. The corresponding image formations are shown in the same colour.}
    \label{fig:example_2eNFW}
\end{figure*}

\end{document}